\documentclass[]{spie}  %>>> use for US letter paper
\usepackage[]{graphicx}

\title{Site characterization studies for the Iranian National Observatory} 

\author{Habib G. Khosroshahi
\skiplinehalf
School of Astronomy, Institute for Research in Fundamental Sciences, Tehran, Iran}

\authorinfo{habib@ipm.ir}
\begin{document} 
  \maketitle 

%%%%%%%%%%%%%%%%%%%%%%%%%%%%%%%%%%%%%%%%%%%%%%%%%%%%%%%%%%%%% 
\begin{abstract}
We report on the Iranian National Observatory (INO) ongoing site
characterization studies for INO 3.4m optical telescope under
development. Iran benefits from high altitude mountains and a
relatively dry climate, thus offer many suitable
sites for optical observations. The site selection (2001-2007) studies
resulted in two promising sites in central Iran, one of which will
host the 3.4m telescope. The studies between 2008 and 2010 aimed at 
detail characterization of the two sites. This involved measurements
of a number of parameters including the wind speed and wind direction,
astronomical seeing, sky brightness and microthermal variations.
\end{abstract}

%>>>> Include a list of keywords after the abstract 

%%%%%%%%%%%%%%%%%%%%%%%%%%%%%%%%%%%%%%%%%%%%%%%%%%%%%%%%%%%%%
\section{INTRODUCTION}
\label{sec:intro}  % \label{} allows reference to this section

The present research and training capabilities in observational
astronomy in Iran can, by no mean, respond to the growing demand due
to the rapid growth in higher education over the past decade. The
existing observational facilities consists of a number of
small telescopes in various university campus observatories generally
used for undergraduate and graduate training. A medium size optical
telescope is thought to be a step to facilitate research in astronomy
and observational cosmology. The geographic location of Iran, 32N
53E, relative dry climate and high altitude mountains, offer suitable
locations for optical telescopes.

Site selection study for a proposed 2-4 meter class telescope started
few years before the INO project received administrative approval. The
study led by S. Nasiri (report in preparation) began by collecting and
analysis of weather data, seismic hazard data, accessibility and shinny day
statistics over central dry regions of the country. A large number of sites 
were identified and inspected. When the number of potential sites, mostly 
scattered around the central desert, was reduced to a manageable number, long
term seeing monitoring has also started and continued for two years on
4 different sites with altitudes between 2500m and 3000m.

%%%%%%%%%%%%%%%%%%%%%%%%%%%%%%%%%%%%%%%%%%%%%%%%%%%%%%%%%%%%%
\section{Site characterization} 

It has been shown that the atmospheric turbulence has a strong
connection to astronomical seeing. In particular the
Fried parameter, $r_0$, which represents the telescope aperture
diameter, for which the diffraction-limited image resolution is equal
to the FWHM of the seeing-limited image is shown to be determined by
refractive index structure constant (Fried 1966) which itself depends
on the temperature structure of the atmosphere (e.g. Marks et al
1996).

Site characterization involved measurement of a number of key site
parameters such as the wind speed and direction, sky brightness,
seeing and microthermal variation profile at the two sites, 
known as Dinava (3000m) and Gargash (3600m). These two sites are 70km apart.
The key objective of the monitoring was to find the best of the two 
sites for the installation of the 3.4m telescope.

\subsection{Wind speed and direction}
Typical weather stations were installed in both sites on 12m masts by
the end of 2008. They allowed the measurement of temperature, wind
speed and direction, barometric pressure and humidity. Wind data
recording was performed every 10 minutes at an 8m height above the
peak. Two years of measurement indicates that both sites shows a
peak wind speed of 4.0-8.0 m/s but despite a 600m higher altitude, the
wind speed in Gargash is generally lower than in Dinava. The west and
south-west are generally the dominant wind directions in both sites. 
This is shown in Fig 1.

   \begin{figure}
   \begin{center}
   \begin{tabular}{c}
   \includegraphics[height=8cm]{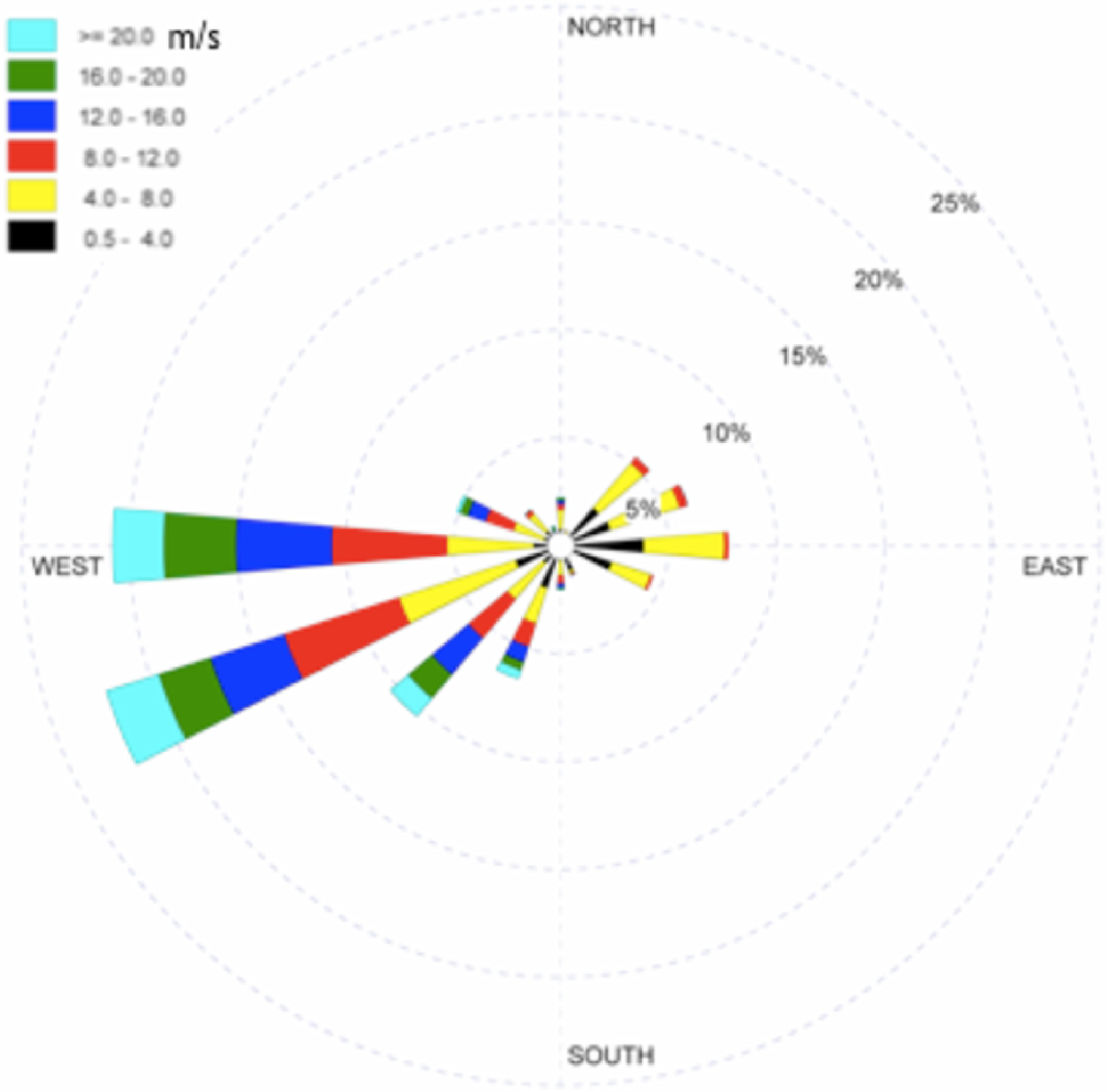} \hspace*{1cm}
   \includegraphics[height=8cm]{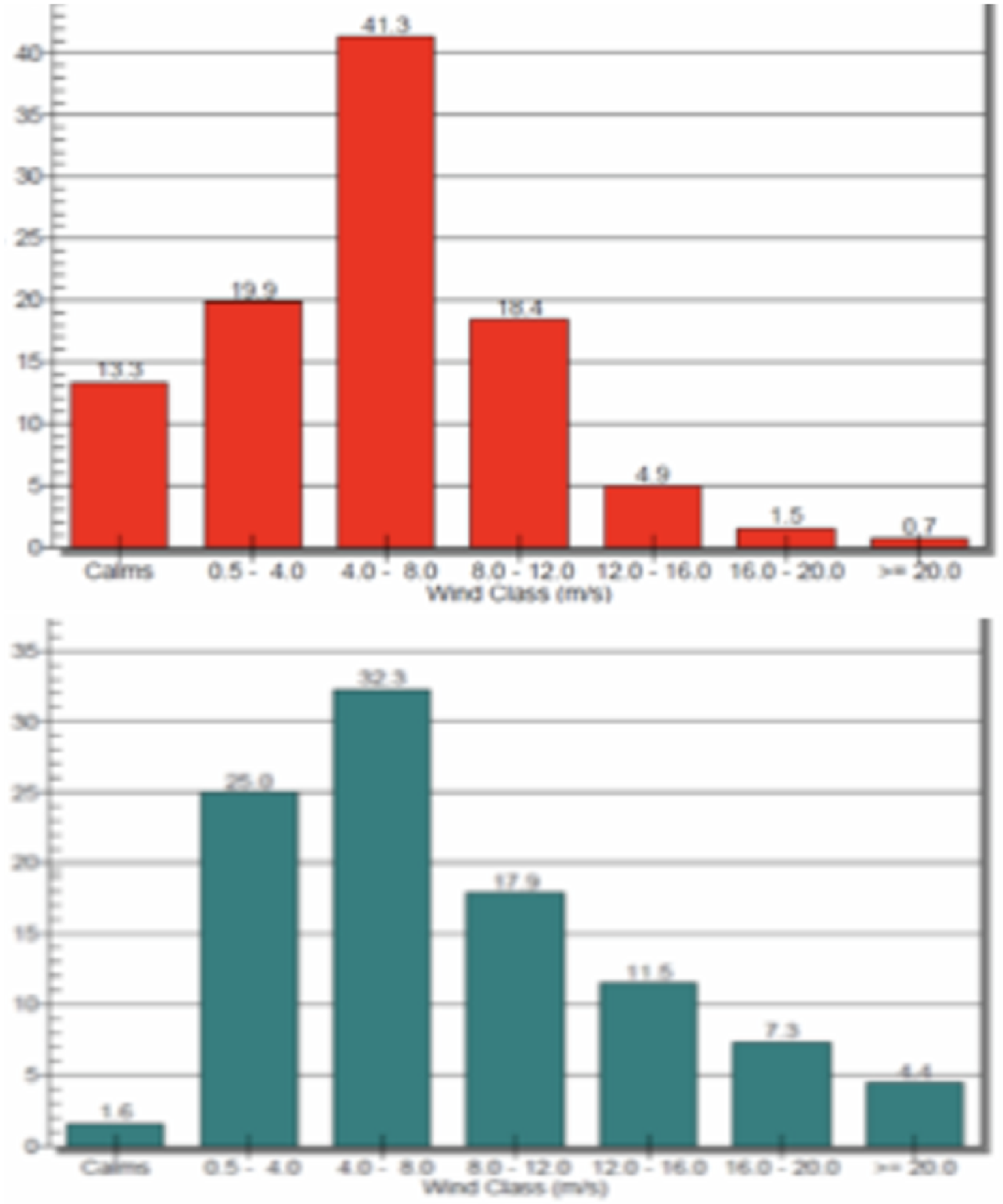}
   \end{tabular}
   \end{center}
   \caption[example] { \label{fig:wind} Left: Gargash site windrose is shown
which clearly indicates a dominant wind direction and its
intensity. The data covers Jan 2009 - Oct 2010. Dinava site shows a
similar windorse. Right: The wind speed histogram is shown for Gargash
(top) and Dinava (bottom) during the same period.}
   \end{figure}

\subsection{Humidity, clear sky and temperature}

Statistically there is about 230 shinny days available for the
region. Monitoring the cloud coverage over two years indicates that
around 45\% clear sky is available annually. This increases to above
70\% between June-Oct.

In about 55\% of the nights the relative humidity remains below
60\%. This increases to over 80\% between May-Oct. There is no
measurable difference between the two sites in relative humidity.

Temperature variation ($T_{max}-T_{min}$) during the night (between
twilights) is generally 3 degrees. The temperature changes at a rate
of about 0.15 ($\pm0.3$) degree celsius per hour between sunset and
midnight.  Dinava site is generally about 5 degrees celsius warmer
than the Gargash site.

\subsection{Seeing measurement}

Seeing is one of the most important parameters describing the
atmospheric turbulence. Seeing measurement was carried out using DIMM
systems (eg. Sarazin \& Roddier 1990, Vernin \& Munoz 1995, Tokovinin
2002) which comprised of Orion Ritchey-chretien 8 inches telescopes,
44 mm apertures with a 122 mm separation, installed on metal pillar
located on a lifted concrete platform providing an altitude of 3.5m
above the ground for the telescopes in both sites. The two DIMM
systems installed in Gargash and Dinava were cross calibrated at Dinava site
using the same configuration. This configuration was kept unchanged
for the period of observations June-Oct 2010. A similar method was
adopted by the site selection team (2004-2006) using 11-inch
telescopes, but on conventional telescope tripod. A comparison of
the measured seeing in Dinava and Gargash is shown in Fig 2.

   \begin{figure}
   \begin{center}
   \begin{tabular}{c}
   \includegraphics[height=12cm,angle=-90]{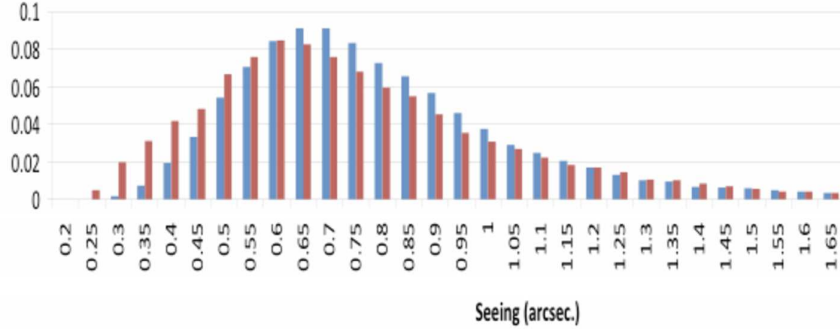}
   \end{tabular}
   \end{center}
   \caption[example] 
   { \label{fig:seeing} 
Seeing distribution compared simultaneously between Gargash (red) and
Dinava (blue) sites in summer 2010 obtained from similar DIMM
systems. The first quartile seeing in Gargash at 3600m is 0.54
($\pm0.04$) arcsec compared to 0.60 ($\pm 0.09$) arcsec. Second
quartile seeing in Gargash is 0.67 arcsec and 0.72 arcsec for Gargash
and Dinava, respectively.}
   \end{figure}

\subsection{Microthermal variation measurement and CFD modeling}
The main aim of the microthermal measurements is to determine the
height of ground layer turbulence which allows an optimization of the
cost-height, driven by desire to located the primary mirror above the
turbulent layer. In case of complex peak topography, multiple
measurements further helps to better constrain the location of the
telescope/enclosure.

As the time-scale of the temperature variation is of the order of
10-100Hz and the amplitude of the variation is of the order of 0.01 of
a degree, the sensitivity of the sensors and the data recording system 
as well as their response time should be adequately set.

We therefore designed a system to deliver $\sim$1 kHz recording frequency
with a few $\times$ 0.001 degree sensitivity using Platinum wire with high
purity and 20 micron diameter. 

   \begin{figure}
   \begin{center}
   \begin{tabular}{c}
   \includegraphics[height=12cm,angle=-90]{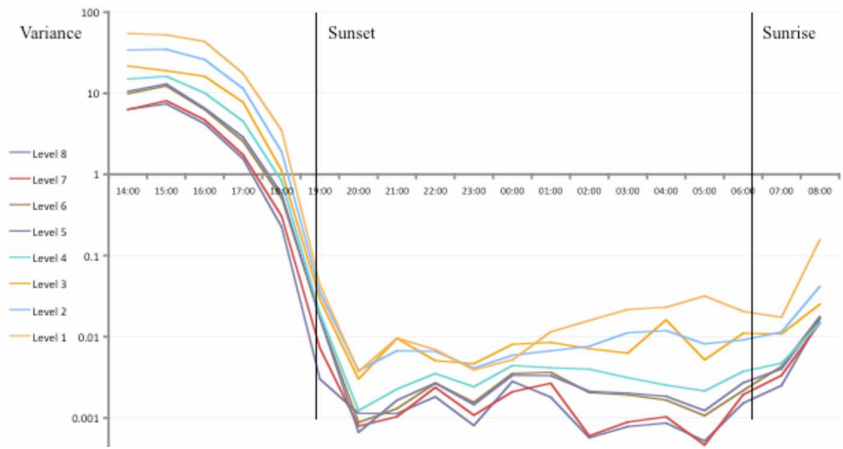}
   \end{tabular}
   \end{center}
   \caption[example] 
   { \label{fig:seeing} 
An example microthermal variation profile for one of the masts in Gargash. The median of the variance in each level is obtained for 6 consecutive days in Sept 2010. The lower levels show larger variance during the day and night relative to the upper levels.}
   \end{figure} 

The microthermal variation measurements were performed in 6 locations (given
the complexity of the peak topography) in Gargash site and 2 locations in
Dinava site simultaneously in September-October 2010. The sensors were
placed at 8 levels with a separation of 1.5m vertically. The horizontal 
separation of the sensors is 2 meters. A quick analysis of the
results show that first mast in the direction of the dominant wind
direction (shown in Fig 3) provides a textbook example of the thermal
variation profile. There is a clear difference in the recorded variance
between the levels which is observed in day and night time. More
detailed analysis of the microthermal measurements is in progress.

We have obtained the topographic map of the peak with resolutions of 1
meter and 5 meters for the upper (-30 meter from the peak) and lower
(-100 meters of the peak) regions of the peak to be able to perform a
Computational Fluid Dynamics (CFD) modeling of the peak under various
wind flow and turbulence conditions. Our initial findings indicate
that the boundary layer is about 15-20 meters from the ground.

\subsection{Sky brightness}

Sky background was measured under photometric conditions in Dinava and
Gargash. We find that Gargash site is about 0.4 magnitude darker than
the Dinava site owing to a larger distance from major cities. The V-band
sky brightness in Dinava and Gargash are 21.6 and 22.0 mag, respectively. A
light pollution control project is being planned to preserve the sites
for astronomical observations. 

\section{Concluding remarks}

Our studies indicate a relative advantage of the Gargash site 
in comparison to Dinava site. Gargash site is found to be darker,
benefitting from a better astronomical seeing and also higher altitude
and therefore less affected by dust. 

%%%%%%%%%%%%%%%%%%%%%%%%%%%%%%%%%%%%%%%%%%%%%%%%%%%%%%%%%%%%%
\acknowledgments
 
The site selection activity was handled by Institute for Advanced
Studies Basic Sciences (led by S. Nasiri) between 2001 and 2007. Site
characterization and monitoring reported here was handled by the
Iranian National Observatory Project team and the Institute for
Research in Fundamental Sciences (IPM). I acknowledge the contribution
of individuals, R. Mansouri, A. Ardeberg, S. Arbabi, A. Haghighat,
A. Behnam, A. Molainejad, A. Roozrokh, A. Danesh, A. Jafarzadeh,
R. Ravani, A. Mirhoseini, B, Afzalifar, F. Ghaderi and site monitoring
teams.

%%%%%%%%%%%%%%%%%%%%%%%%%%%%%%%%%%%%%%%%%%%%%%%%%%%%%%%%%%%%%
%%%%% References %%%%%
\section*{REFERENCES}

Fried D.L., 1966, J. Opt. Soc. Am. 56, 1372\\
Marks, R.D., Vernin J., Azouit M., Briggs J.W., Burton M.G., Ashley M.C.B and Manigault J.F., 1996, Astron. Astrophys. Suppl. Ser. 118, 385\\
Sarazin, M.; Roddier, F., 1990, A\&A, 227, 294\\
Tokovinin, A., 2002, PASP, 114, 1156\\
Vernin, J., Munoz-Tunon, C., 1995, PASP, 107, 265\\ 	

%\bibliography{report}   %>>>> bibliography data in report.bib
%\bibliographystyle{spiebib}   %>>>> makes bibtex use spiebib.bst

\end{document}